\documentclass[aps,prd,superscriptaddress,nofootinbib,eqsecnum,twocolumn]{revtex4-1}

\usepackage{amsfonts}
\usepackage{amsmath}
\usepackage{amssymb}
\usepackage{tikz}
\usepackage{bm}
\usepackage{dcolumn}
\usepackage{graphicx}   
\usepackage[latin1]{inputenc}
\usepackage{latexsym}
\usepackage{rotating}
\usepackage{hyperref}
\usepackage{subfigure}
\usepackage{color}
\usepackage{changes}

\begin{document}

\title{Gravitational Particle  Production  and the Validity of
  Effective Descriptions in Loop Quantum Cosmology}
  
  \author{G.~S.~Vicente}\email{gustavo@fat.uerj.br}
\affiliation{Faculdade de Tecnologia, Universidade do Estado do
  Rio de Janeiro, 27537-000 Resende, RJ, Brazil}

\author{Rudnei O. Ramos}
\email{rudnei@uerj.br}
\affiliation{Departamento de F\'{\i}sica Te\'orica, Universidade do
Estado do Rio de Janeiro, 20550-013 Rio de Janeiro, RJ, Brazil}

\author{L.~L.~Graef}
\email{leilagraef@id.uff.br}
\affiliation{Instituto de F\'{\i}sica, Universidade Federal Fluminense,
24210-346 Niter\'oi, RJ, Brazil}

\begin{abstract}

The effective approach in Loop Quantum Cosmology (LQC) has provided
means to obtain predictions for  observable quantities in LQC
models. While an effective dynamics in LQC has been extensively
considered in  different scenarios, a robust demonstration of the
validity of effective descriptions for the perturbative level still
requires further attention. The consistency of the description adopted
in most approaches requires the assumption of a test field
approximation, which is  limited to the cases in which  the
backreaction of the particles gravitationally produced can be safely
neglected. Within the framework of LQC, some of the main approaches to
quantize the linear perturbations are the dressed metric, the  hybrid
approaches and the closed/deformed algebra approach. Here, we analyze
the consistency of the test field assumption in these frameworks by computing the  energy density stored
in the particles gravitationally produced compared to the background
energy density. This analysis ultimately provides us with a
consistency test of the effective descriptions of LQC. 
 
\end{abstract}

\maketitle 

\section{Introduction} 

In the early 1980s the inflationary scenario brought new perspectives for connection between  fundamental physics with experiment.
Inflation was the first paradigm to make concrete 
predictions for the structure of the large-scale Universe  based
on causal physics~\cite{Brandenberger:1999sw}.  Many decades later,
with the improvement on  experiments aiming to  accurate measure the
Cosmic Radiation Background (CMB), several inflationary scenarios
still show good agreement with
data~\cite{Akrami:2018odb,Aghanim:2018eyx}. However, as it is well
known,  many inflationary scenarios require the fields to be 
 very homogeneous initially or start with fine tunned  initial
conditions. This leaves inflation at a crossroads, since General
Relativity (GR) inevitably implies an initial singularity, where it
is not clear how one should impose the initial conditions.  Inflation is very
sensitive to Planck-scale physics~\cite{Baumann:2009ds}.  The
assumption adopted in inflationary cosmology that ``the spacetime can
be treated classically" is clearly questionable. The well-known
successes of inflation motivate the community to search for means to
past complete this cosmological scenario with a more fundamental and
consistent quantum gravity theory in the ultraviolet (UV) scale.

Cosmological spacetimes have
the  advantage  of simplicity for  being  highly  symmetric,  since  
homogeneity  reduces  to a finite number, the infinite number of degrees that one would have otherwise. This favors the development of spacetime quantization schemes. In particular, it worth mentioning the recent
 progresses on the quantization of 
cosmological spacetimes using the approach of loop quantum gravity
(LQG), a  nonperturbative  quantum gravity theory which has opened new
avenues to explore Planck scale physics.   The  reduced  version of LQG is Loop  Quantum  Cosmology
(LQC)~\cite{Bojowald:2001xe,Ashtekar:2006rx,Ashtekar:2006uz,Ashtekar:2006wn,Ashtekar:2007em,Singh:2009mz,Ashtekar:2006es,Szulc:2006ep,Ashtekar:2011ni,Barrau:2013ula,Ashtekar:2009mb}, an approach which uses the symmetries
considered in cosmology.  
Besides allowing for the construction of  non-singular early Universe models, the increasing progress on the analysis of
cosmological fluctuations in LQC has bridged quantum gravity with
cosmological observations~\cite{Agullo:2016tjh}.

In the framework of LQC, the background evolution can be divided into
two classes: The kinetic dominated bounce and the potential dominated
bounce. A bounce which is dominated by potential energy would  either fail in producing sufficient slow-roll inflation or lead to a too large amount of  expansion~\cite{Ashtekar:2015dja}. In the latter case,
all the new physics is washed out, and no signal from the quantum
regime is present.  In the case of kinetic energy initial  domination,
the early evolution can always be divided into three phases after the
contraction: the bouncing phase (with equation of state $\omega=1$),
the transition phase ($-1<\omega<1$) and inflation ($\omega \simeq
-1$). The presence of these three stages is universal in the kinetic
dominated case and does not depend on the form of the inflaton
potential.

Unlike the general evolution of the background (zero-modes), the
linear perturbations depend on the methods used to quantize
them. Within the framework of LQC, there are mainly four different
approaches: the dressed metric~\cite{Agullo:2012sh,Benetti:2019kgw},
the closed/deformed algebra~\cite{Bojowald:2008gz,Cailleteau:2011kr},
the hybrid~\cite{FernandezMendez:2012vi} and separated universe
approach~\cite{Wilson-Ewing:2015sfx,Wilson-Ewing:2016yan}.

In the dressed metric approach, the perturbative degrees of freedom are
quantized using the {}Fock quantization procedure while the background metric is quantized by
the loop method.  The quantum dynamics of the perturbations can be
described by a quantum field evolving in a dressed background
 in the cases in which the
energy density of the perturbations are small  compared to the
Planck energy. In the dressed metric approach, and also in the other
approaches considered in this work,  if a 
potential is added to the dominant field, the dynamics becomes
too complicated and in this case an effective field theory needs to be adopted.  

In the hybrid quantization approach, the background and the perturbed
degrees of freedom are also treated differently, since a LQG-like
quantization of the background is performed along with a Fock-like
quantization of the
perturbations~\cite{Gomar:2014faa,Martin-Benito:2008eza,MenaMarugan:2009dp,Garay:2010sk,Martin-Benito:2010dge,Brizuela:2009nk}. As
a result of the non-homogeneous degrees of freedom being not loop but
{}Fock quantized, the kinematic Hilbert space is a tensor product of
the individual Hilbert space for each sector, that is,
$\mathcal{H}_{kin}= \mathcal{H}_{kin}^{grav} \otimes
\mathcal{H}_{kin}^{matt} \otimes \mathcal{F}$~\cite{Li:2020mfi}.
While the background geometry is loop quantized, the zero-mode of the
scalar field is quantized in the standard Schr\"{o}dinger representation,
and the non-homogeneous perturbations are {}Fock quantized. Similar to
the dressed metric approach, for sharply peaked semiclassical
background states, there  exists an effective description of the
quantum dynamics,  which greatly simplifies the dynamical
equations~\footnote{For a different treatment on the hybrid approach,
  which does not consider effective background equations, see, for
  example, Ref.~\cite{Gomar:2015oea}.}. Despite the hybrid approach also
providing effective perturbative equations  that are similar to
the dressed approach, sharing several similarities with the latter,
 they incorporate differently quantum gravity
corrections. This leads  to a few 
differences~\cite{Gomar:2017yww}.  In Ref.~\cite{Li:2020mfi} a robust evidence was
provided of differences in predictions between
dressed and hybrid approaches due to the  respective underlying
constructions in the context of the modified LQC-I model
(mLQC-I)~\cite{Li:2019ipm,Li:2018fco}. Although it has been shown
that, in the effective description, these approaches  do not lead to
significant differences in the observable parts of the spectrum, the
behavior of the non-observable part of the CMB spectrum deserves
investigation. As shown in Ref.~\cite{Graef:2020qwe}, for the case of
the dressed metric approach, the behavior of the  non-observable range
of mode frequencies would imply, if it were not  for invalidating the
effective description of the approach, in a
pre-inflationary phase dominated by radiation, which would delay and
shorten the inflationary phase. 

Another important framework is the closed/deformed algebra
approach~\cite{Barrau:2017tcd,Li:2018vzr}, which considers effective constraints coming from the quantum
corrections. In this approach the effective constraint algebra must be closed after the
quantum corrections are considered. In this approach one encounters  the problem that the 
signature of space-time effectively   change from Lorentzian to
Euclidean at  high
curvatures ~\cite{Li:2018vzr,Schander:2015eja}. The transition point
between Lorentzian and Euclidean space-time implies a "state of
silence", characterized by a vanishing speed of sound, which 
can be interpreted as due to a decoupling of different space
points~\cite{Mielczarek:2014kea}. This effect comes from the necessity to have a
closed/deformed algebra of quantum corrected effective constraints
when  including  holonomy corrections from loop quantum gravity. 

Finally, it is important to also mention the separate universe
approach~\cite{Wilson-Ewing:2015sfx,Wilson-Ewing:2016yan}. This approach considers a spacetime with small perturbations  discretized in a
lattice. At each cell, which is considered  to be homogeneous and noninteracting, loop quantization is applied. The dynamics of cosmological perturbations can  be approximated by
effective equations whenever we have a sharply peaked wave function in each cell. We also do not consider this approach  since its results are only applicable to infrared modes.

An important open question in all of the above approaches concerns the
choice of initial conditions for the evolution of the
non-homogeneities. Usually, in the traditional scenario, the initial
data are set at the beginning  of inflation or during the slow-roll
phase. However, the situation is very different in LQC, since it is
not clear  where to set the initial conditions and whether a vacuum
state can be defined at those points. This is one of the most
important conceptual questions to be understood in such scenarios.  As
shown in
Refs.~\cite{Li:2018vzr,Zhu:2017jew,Linsefors:2013cd,Martineau:2017sti,Bolliet:2017czc,Ashtekar:2011rm,Graef:2018ulg,Barboza:2020jux,Barboza:2022hng},
in addition to the differences at the perturbative level, different
choices of initial conditions can severely  affect  the duration of
inflation in LQC.

While in the dressed metric and hybrid approaches the initial
conditions are set either in the contracting phase or in the bounce,
in the closed/deformed algebra approach the initial conditions must be
set at the silent point, soon after the bounce. This state is the
beginning of the Lorentzian phase in this approach, and, in some
sense, the beginning of time~\cite{Mielczarek:2014kea}. As discussed
in Ref.~\cite{Li:2018vzr}, this is the only known initial condition
that can lead to a spectrum compatible with the observations in this
approach. On the other hand, in the dressed metric and hybrid
approaches, both the usual Bunch-Davis vacuum and adiabatic-like
initial conditions for the perturbations at the bounce, or at the
contracting phase, have been considered. 

As mentioned before, most of the approaches in LQC rely on effective
descriptions to provide  more treatable  equations for the
non-homogeneous sector. For such an effective description to be valid,
the  backreaction of the modes gravitationally produced must be
negligible. However, according to  the Parker Gravitational Particle
Production (GPP)  mechanism~\cite{Parker:1968mv,Parker:1969au}, a test
scalar field $\chi$, evolving from a pre-bounce Minkowski vacuum state
to a post-bounce different Minkowski vacuum state, will develop a final
state containing $\chi$ particles, similarly to linear cosmological
perturbations on the cosmological background.  This effect is usually
negligible in inflationary phases, but in a  bounce phase the
situation can be
different~\cite{Quintin:2014oea,Tavakoli:2014mra,Haro:2015zda,Celani:2016cwm,Scardua:2018omf,Zago:2018huk,Hipolito-Ricaldi:2016kqq,Graef:2017nyv,Tavakoli:2014mra,Haro:2015zda}.
In the dressed metric approach, the gravitational particle production
was calculated in Ref.~\cite{Graef:2020qwe}. In that work it was shown
that the relativistic particles gravitationally produced during the
bounce would come to dominate the energy density of the Universe
before inflation, which invalidates the test field approximation
required in this approach. Motivated by this result, we revisit the
particle production in the dressed metric approach and we also
investigate  whether the test field approximation is valid in the
closed/deformed algebra and in the hybrid approaches. We proceed  by
comparing the energy density stored in particles gravitationally
produced with the background energy density  in each approach. 

This paper is organized as follows.  In Sec.~\ref{sec2}, we describe
the background dynamics of the LQC model with a kinetic dominated
bounce.  In Sec.~\ref{sec3}, we present the dynamics of the
perturbative modes in the dressed metric, hybrid and deformed/closed
algebra  approaches and introduce the  GPP mechanism.  In
Sec.~\ref{sec4}, we present the  results for the energy density of the
particles produced in each case. {}Finally, our concluding remarks are
presented in Sec.~\ref{sec5}.

\section{Background Model}
\label{sec2}

We consider LQC as  the quantum background scenario, which provides GR
in the classical regime and quantum corrected GR equations in the
Planck regime.  The scale factor $a$ arises from the definition of a
 fiducial fixed cubic cell, with a volume described by $v = {\cal V}_0
a^{3} m_{\rm Pl}^2/(2\pi\gamma)$, ${\cal V}_0$ being the comoving
volume of a cell in LQC and $\gamma$  the Barbero-Immirzi parameter
 whose value we are going to consider to be $\gamma\simeq
0.2375$, according to black hole entropy
calculations~\cite{Meissner:2004ju}. Above $G$ is the Newtonian
constant of gravitation and the Planck mass is $m_{\rm Pl}\equiv 1/\sqrt{G} =
1.22 \times 10^{19}\,$GeV.  The variable 
$b$ denotes the conjugate momentum to $v$ and it is given by $b=-4\pi\gamma P_{(a)}/(3 a^2 {\cal V}_0 m_{\rm
  Pl}^2)$, where $P_{(a)}$ is the momentum conjugate  to the scale
factor.  

The quantum {}Friedmann equation, obtained by solving the effective LQC 
equations, reads~\cite{Ashtekar:2011rm}:
\begin{equation}
\frac{1}{9}\left(\frac{\dot{v}}{v}\right)^{2} \equiv H^2
=\frac{\sin^2(2 \lambda b)}{4 \gamma^2 \lambda^2} = \frac{8\pi}{3
  m_{\rm Pl}^2} \rho \left(1- \frac{\rho}{\rho_{\rm c}} \right),
\label{hubble}
\end{equation}
where $\lambda=(48\pi^2\gamma^2/m_{\rm Pl}^4)^{1/4}$ and $b$ lies in the interval 
$(0, \pi/\lambda)$. Above, $\rho$ accounts for the energy density, $\rho_c = 3 m_{\rm
  Pl}^2/(8\pi\gamma^2 \lambda^2)\approx 0.41m_\mathrm{Pl}^4$ represents the
critical density  and the dot represents derivative with respect
to the cosmological time.  The energy density $\rho$ is connected to the 
variable $b$ through the relation $\rho = 3 m_{\rm Pl}^2 \sin^2(\lambda
b)/(8\pi\gamma^2 \lambda^2)$. {}For an energy density much smaller than the critical density  we
reobtain GR as expected in the classical limit.  Due to quantum effects, the
singularity is not present in this framework and a bounce phase is obtained
when the energy density has a value close to the critical density. After the
bounce, the Universe transits into a decelerated expansion phase with
a subsequent inflationary phase.

The inflaton field can be considered as behaving as a 
fluid with equation of state  $p=\omega \rho$. The
solution for the scale factor in LQC for single fluid is given by (see, e.g.,
Ref.~\cite{Wilson-Ewing:2013bla})
\begin{eqnarray}\label{aLQC}
a(t)=a_B \left[1+\frac{\gamma_B(1+\omega)^2}{4} \left(\frac{t}{t_{\rm
      Pl}}\right)^2\right]^{\frac{1}{3(1+\omega)}},
\end{eqnarray}
where $\gamma_{B}\equiv 24\pi\rho_{c}/m_\mathrm{Pl}^{4}\simeq 30.9$
while $t_\mathrm{Pl}\equiv 1/m_\mathrm{Pl}$ is the Planck time.  The
evolution until the end of inflation can be divided mainly into a contracting
phase, a bounce phase and the classical slow-roll phase. 

We consider a cosmological background dominated
by the inflaton field with  equation of motion,
\begin{eqnarray}\label{inflaton}
\ddot \phi + 3H \dot \phi + V_{,\phi}=0,
\end{eqnarray}
where $V(\phi)$ is the potential energy of the field. We  consider in this setup also an extra scalar field denoted by $\chi$, that
 couples to $\phi$ and couples gravitationally to the Standard Model particles
, which behaves  as a {\it spectator}
field~\cite{Agullo:2012sh,Agullo:2013ai,Ashtekar:2009mb,Agullo:2012fc}.
This field will be produced gravitationally in the bounce phase and it
is assumed to be dynamically relevant only in the post-bounce
evolution, particularly in the pre-inflationary dynamics, where it
might behave as radiation.If we consider that $\chi$ has a 
 very small mass in comparison with the post bounce $H$ parameter, then $\chi$ will behave  as radiation in the
pre-inflationary phase.  The analysis of the GPP of a spectator scalar
field  has already been considered in the dressed metric
approach~\cite{Agullo:2012sh,Agullo:2013ai,Ashtekar:2009mb,Agullo:2012fc}
and in the hybrid approach~\cite{Prokopec:2017ldn}.  The equation of
motion for the  scalar perturbations as well as the scalar field equations have the same form as  in  GR. Later we are going to further analyze the particle production associated to the field
 $\chi$. Here we will follow an analysis
similar to the one used in Ref.~\cite{Graef:2020qwe}, which was
originally applied to the dressed metric approach.

Below we
summarize the results of Ref.~\cite{Bolliet:2015bka}, where a full
computation of the background dynamics in each phase can be found. As
an example, we consider in the following the chaotic model for the
inflaton field. However, as we will see,   the results for particle
production in the dressed and hybrid approaches do not depend on the
choice of the potential for the inflaton field, as the GPP happens
mainly during the kinetic energy dominated bounce phase. However, we
will see that, to obtain results for the closed algebra approach, they
will explicitly involve the background dynamics after the bounce
phase.

{}For later reference, let us briefly review below the background
dynamics in LQC. We consider the chaotic quadratic inflaton potential
$V(\phi)=m^2 \phi^2/2$ as an example, although the overall description
is not expected to change significantly for other forms of potentials.

\paragraph{\bf Contracting Phase:}

The scale factor in the classical contracting phase, written in terms
of the conformal time $\eta$ ($dt=a d\eta$),  follows the expression
\begin{equation}\label{bouncephase}
a(\eta)=\lambda_{0} \eta^{2}, \; \; \; \; {\rm with} \; \; \; \;
\lambda_{0}=\frac{a_{in}H_{in}^{2}}{4},
\end{equation}
where $a_{in}$ and $H_{in}$ are the initial values for the scale
factor and for the Hubble parameter, respectively.  

There are two time scales in our original system of equations. One is
given by $1/m$, associated with the classical evolution of the
inflaton field. The other one is $1/\sqrt{G\rho_{c}}$, which is
associated with the quantum regime. The ratio between these two
timescales is defined by the quantity $\Gamma$:
\begin{equation}\label{Gammadef}
\Gamma = \frac{m}{\sqrt{24\pi G \rho_{c}}},
\end{equation}
where $\Gamma \ll 1$. Here, we assume $m=10^{-6}m_{\rm Pl}$, as
suggested by the observations. Since $\rho_{c}=0.41m_{\rm Pl}^4$, this
leads to $\Gamma \sim 2 \times 10^{-7}$.  By also defining
\begin{equation}
x(t) \equiv \frac{m\phi}{\sqrt{2\rho_{c}}}, \;  \; \; \;  \;  y(t)
\equiv \frac{\dot{\phi}}{\sqrt{2\rho_{c}}},
\end{equation}
in the classical contracting phase, $x$ and $y$ can be
expressed as
\begin{eqnarray}
x(t)&=&\sqrt{\frac{\rho(t)}{\rho_{c}}} \sin(mt+\theta_{0}),
\\ y(t)&=&\sqrt{\frac{\rho(t)}{\rho_{c}}} \cos(mt+\theta_{0}).
\end{eqnarray}
When  $H \approx -m/3$, the  term proportional to $H$ in
Eq.~\eqref{inflaton} becomes dominant. It can be considered as  the
end of the pre-bounce contracting phase and the start of the bouncing
phase. We denote the density at the end of the contracting phase by
$\rho_A$, which is given by $\rho_{A}=\Gamma^{2} \rho_{c}$, so that
before the bounce phase starts, there are still no significant quantum
effects. 

\paragraph{\bf Bounce Phase:}

We can define the starting of the bounce phase when $\rho=\rho_{A}$.
At this time, the quantities $x$ and $y$ can be written as
\begin{eqnarray}
x_{A}= \Gamma \sin \theta_{A}, \; \; \; \;  y_{A}=\Gamma
\cos\theta_{A}.
\label{xy}
\end{eqnarray}
The inflaton field kinetic energy dominates in the bounce phase and
this phase is then like stiff matter, i.e., like a fluid with equation
of state $\omega\approx 1$. {}From Eq.~(\ref{xy}), one can see that we
must have $cos \; \theta_{A} \sim 1$ due to the kinetic energy
domination in this phase.  In particular, with ($\omega\approx 1$)
from Eq.~\eqref{aLQC}, the scale factor reads
\begin{eqnarray}\label{aLQCstiff}
a(t)=a_B \left(1+\gamma_B \frac{t^2}{t_{\rm
    Pl}^2}\right)^{\frac{1}{6}}.
\end{eqnarray}
The evolution of the inflaton field in this phase is described by
\begin{eqnarray}\label{phiLQCstiff}
\phi(t)=\phi_{\rm B}\pm \frac{m_{\rm pl}}{2\sqrt{3\pi}}{\rm
  arcsinh}\left(\sqrt{\gamma_B}\frac{t}{t_{\rm pl}}\right),
\end{eqnarray}
where the plus sign applies when $\dot \phi >0$ and the minus sign for
$\dot \phi < 0$.  The inflaton amplitude at the bounce, $\phi_{B}$,
expressed in terms of the variable $x_{B}$, can be written as
\begin{equation}
 x_{B}=x_{A} - \epsilon \Gamma \;  \ln\left(\frac{1}{2}\Gamma \;
 \cos\theta_{A}\right),
\end{equation}
where $\epsilon \equiv {\rm sgn}(\cos \;  \theta_{A})$.

\paragraph{\bf Slow-Roll Phase:}
 
 In the starting of  the slow-roll phase,  a time we  denote by
 $t_{SR}$, the energy density is $\rho \ll \rho_c$ and
 the Universe is already classical. The time of the beginning of this
 phase can be determined by solving $\dot{\rho}(t_{SR})=0$. Having this condition we can obtain the relation  $t_{SR}=t_{B}+f/m$, in which 
 $f$ can be written in terms of  $W$ (the Lambert function), which is the 
 solution of the equation $z=W(z)e^{W(z)}$. The function $f$ is given by 
\begin{equation}\label{fdef}
f \equiv \sqrt{\frac{2}{W(z)}} \quad,\quad {\rm with} \; \; \; \;  z=
\frac{8}{\Gamma^{2}} \exp \left(\frac{2|x_{B}|}{\Gamma}\right).     
\end{equation}
{}For $\cos \; \theta_{A} \sim 1$ and $\Gamma=2 \times 10^{-7}$ as we
are considering, we have that $f \sim 0.18$.

 At the time $t_{SR}$ we have that~\cite{Bolliet:2015bka}
 \begin{equation}
x_{SR}=x_{A} - 2 \epsilon \Gamma \ln \left(\frac{1}{2} \Gamma
\sqrt{\frac{|\cos \theta_{A}|}{f}}\right).     
 \end{equation}
Shortly after $t_{SR}$, one has that $y_{SR}\equiv -\epsilon
 \Gamma$ and a very small slow-roll parameter is achieved, as
 expected, and it is given by
\begin{equation}
\epsilon_{H} = 3\left|\frac{\Gamma}{x_{SR}}\right|^{2},
\end{equation}
 which for the values of $\Gamma$ and $\cos \theta_{A}$ that we are
 using assumes the value $\epsilon_{H} \sim 0.003$.
 
The Hubble parameter in this phase is given by
\begin{equation}\label{HSR}
H(t) = H_{SR}\left|1 - \epsilon
\frac{\Gamma}{x_{SR}}m(t-t_{SR})\right|,
\end{equation}
where $H_{SR}=\sqrt{8\pi G \rho_{c}/3}|x_{SR}|$ (and
$a_{SR}=a_{B}\Gamma^{-1/3}$).

\section{Solving for the quantum field modes in LQC}
\label{sec3}

We are interested in how the particle production of the spectator scalar field $\chi$
could change the  LQC pre-inflationary phase. 
In this section we will describe
the mechanism of GPP for each approach within the framework of LQC.
 
The quantum fields are described using the standard procedure for
classical spacetimes, but using techniques from LQG to incorporate
quantum gravity effects~\cite{Agullo:2012fc}, which are suitable to
treat curvature and matter densities at the Planck scale.   We first
introduce each approach of LQC we are considering in this paper,
highlighting the equations of motion for the {}Fourier modes of the
spectator fields.   In the sequence, we introduce the general details
about the Parker mechanism for these fields, which has irrelevant interactions with the other components of the Universe,
except for gravity.

One very important aspect is that in order to have a solvable model,
the usual procedure is to assume effective equations based on the
supposition that quantum corrections due to fluctuations are small
enough so that they have negligible influence on the evolution of
expectation values. Including a significant  backreaction would result
that the evolution becomes more quantum, i.e.,  more dependent on how
the quantum variables behave. The states can then be deformed  from a Gaussian initial distribution. The backreaction
results in  a change on the quantum state shape, and  this then affects
the motion of its expectation values. This effect is important for the
long-term evolution of cosmology. Therefore, in order for the usual
quantization scheme to be valid and to obtain a consistent
solution in the  effective description, we must assure that the
energy density of perturbations is negligible compared to the background energy density ~\cite{Agullo:2013ai}
  during the whole
evolution. Either the  backreaction is ignored in the effective
equations, or we need to consider a complete quantum gravity theory (for a
further discussion on this aspect, see for example
Ref.~\cite{Bojowald:2008pu}).

Regarding the evolution of the spectator field, we work directly in
terms of its {}Fourier $k$-modes $\chi_k$.  The {}Fourier expansion of
the field $\chi$ in terms of the $k$-modes, in conformal time, reads
\begin{eqnarray}
\!\!\!\!\!\!\!\!\!\!\!\!\chi({\bf x},\eta)\! =\!\!\int
\frac{d^3k}{(2\pi)^{3/2}}\!\left[\chi_{k}(\eta)\,a_{\bf k}  e^{-i {\bf
      k}. {\bf x}}\!+\!\chi_{k}^*(\eta)\,a_{\bf k}^\dagger e^{i {\bf
      k}. {\bf x}}\right], 
\end{eqnarray}
where $a_{\bf k}$ and $a_{\bf k}^\dagger$ are the annihilation and
creation operators, respectively, that satisfy the canonical
commutation relation.  In the following, we introduce the dynamics for
the {}Fourier modes of the spectator field $\chi_k$ in each  LQC
approach that we will be considering in this paper. This will later
provide us with means to verify the test field supposition,
$\rho_\mathrm{pert}/\rho_\mathrm{bg}\ll1$.

\subsection{Dressed Metric Approach}

The effective equation of motion for the spectator field $k$-modes in
the dressed metric approach reads~\cite{Zhu:2016dkn}:
\begin{equation}\label{mudressed}
\chi_{k}''({\eta}) + \left[k^{2} - \frac{{a}''(\eta)}{{a(\eta)}} +
  {U}_d({\eta})\right]\chi_{k}({\eta})=0,
\end{equation}
where $a''/a$ is given by~\cite{ElizagaNavascues:2017avq}
\begin{equation}\label{dda}
\frac{a''}{a} =  \frac{4\pi}{3m_{\rm Pl}^2}a^{2}\left[ \rho
  \left(1+2\frac{\rho}{\rho_c}\right) -  3p
  \left(1-2\frac{\rho}{\rho_c}\right) \right],
\end{equation}
where $p$ is the pressure density, prime here indicates derivative with
respect to conformal time and $U_d(\eta)$ in Eq.~(\ref{mudressed})
is given by $U_d(\eta)=a^{2}(\it{f}^{2}V(\phi) +
2\it{f}V_{,\phi}(\phi)+V_{,\phi \phi}(\phi))$ is the effective
potential, with $\it{f} \equiv \sqrt{2\pi
  G}({\phi}'/a)/\sqrt{\rho}$. In this approach, $U(\eta)$, $a$ and 
$\eta$ refer to the background state  quantum expectation values,  $\Psi_{0}(a,\phi)$. {}In the case of background states sharply peaked, as  often considered, we can approximate the dressed effective quantities
 by their peaked values $U_d(\eta)$, $a$ and $\eta$.
. 

The effective potential $U_d$ can be shown~\cite{Zhu:2017jew} to be
negligible during the whole bounce and transition phases. Therefore the equation of motion in these regimes can be written as
\begin{eqnarray}\label{muk}
\chi_k''(\eta) + \left[k^2 -
  \frac{a''(\eta)}{a(\eta)}\right]\chi_k(\eta)=0,
\end{eqnarray}
where $-a''(\eta)/a(\eta)$ corresponds to an effective square mass
term for the modes.

From the scale factor, Eq.~\eqref{aLQCstiff}, we can define the
characteristic momentum scale at the bounce,
$k_B=\sqrt{a''(t)/a(t)}|_{t=t_B}=\sqrt{\gamma_{B}/3}\, a_{B} m_{Pl}$.
The quantity $k_B$ plays the role of the characteristic energy scale at the bounce
in the dressed approach.  It is also important to define the quantity $\lambda =\sqrt{a(t)/a''(t)}$ which is the characteristic
 scale that  plays a role
similar to  the comoving Hubble radius. As it is well known, the modes with $k \gg
k_B$ are  oscillatory since they are inside the effective radius. On the other hand,
modes with $k\approx k_B$ are  inside the effective radius during the contracting
phase and then exiting $\lambda$ during the bounce phase. After that they enter  again the effective radius in 
the transition phase.

\subsection{Hybrid Approach}

The effective equation of motion for the spectator field modes in the
hybrid approach is given
by~\cite{FernandezMendez:2012vi,ElizagaNavascues:2017avq}
\begin{equation}\label{muhybrid}
\chi_{k}''({\eta}) + \left[k^{2} - \frac{4\pi}{3m_{\rm
      Pl}^2}a^{2}\left( \rho   -  3p  \right) +
  \mathcal{U}_h({\eta})\right]\chi_{k}({\eta})=0,
\end{equation}
where the variables are the same ones described in the dressed metric
approach, but now $\mathcal{U}_h(\eta) =  a^2\left( V_{,\phi\phi} +
48\pi G V + 6 a'\phi'V_{,\phi}/(a^3\rho) - 48\pi G V^2/\rho \right)$.
In the bounce phase, where the gravitational particle production is
more relevant, and up to the transition phase, the kinetic energy of
the scalar field dominates the energy content of the Universe.  In
these regimes, we can neglect the contribution of
$\mathcal{U}_h(\eta)$ such that Eq.~(\ref{muhybrid}) becomes
\begin{eqnarray}
\chi_k'' + \left[k^2-\frac{4\pi}{3m_{\rm Pl}^2}a^{2}\left( \rho   -
  3p  \right)\right]\chi_k=0,
\label{chihybrid}
\end{eqnarray}
with $-4\pi a^{2}\left(\rho-3p \right)/(3m_{\rm Pl}^2)$ corresponding
to the effective square mass term for the modes in the hybrid case.
Also, in the transition phase, the energy density drops down to about
$10^{-12}\rho_c$~\cite{Wu:2018sbr}, $a''/a$ in Eq.~(\ref{dda})
reduces to $4\pi a^{2}\left(\rho-3p \right)/(3m_{\rm Pl}^2)$ and we
recover the standard expression of the {}Fourier modes, with the
effective mass term as in Eq.~(\ref{chihybrid}). 

Analogously to the dressed metric approach, we can also define a
characteristic momentum scale in the hybrid approach, which reads
$k_H=k_B/\sqrt{3}$.  The modes behave similarly to the dressed metric
approach, but now with respect to the characteristic momentum $k_H$,
which is subtly different from $k_B$.

We can now draw a parallel between the hybrid and the dressed metric
approaches regarding the impact of its quantization strategies in the
evolution equations for the modes~\cite{ElizagaNavascues:2017avq}.
The are two main differences in the evolution equations.  {}Firstly,
the effective potential $U(\eta)$ is different in each
approach. However, since in the scenarios considered here these
potentials can always be neglected in the relevant moments for GPP,
it does not affect our results.  Secondly, and most important, the
effective mass term is different throughout the evolution in each
approach. 

These differences are due to the proper treatment of the phase space of the
perturbed {}Friedmann-Lema\^{\i}tre-Robertson-Walker cosmologies in
each formalism.  In the hybrid approach the standard procedure is to treat  the whole phase space 
as a symplectic manifold. The effective mass term is thus expressed in
terms of canonical variables,  and the expectation value of the operator
which represents such a canonical expression is then evaluated by using the effective dynamics in LQC. On the other hand, in the
dressed metric case, there is no such global canonical symplectic
structure on the truncated phase space and therefore the effective mass is afterward  evaluated on the  LQC effective solutions.
In Ref.~\cite{ElizagaNavascues:2017avq}, for
example, the difference between the dressed metric and hybrid approaches is
explained in detail.

Despite the differences between the two approaches, the procedure for
obtaining the Bogoliubov coefficients in both approaches, which is
relevant for the GPP, is basically the same.  In both approaches one
can realize that the equation of motion for $\chi_k$ is 
analogous to a  Schr\"{o}dinger-type equation having an 
effective mass term in Eq.~(\ref{muk}) and in Eq.~(\ref{chihybrid})
which acts as  a potential, behaving effectively as a barrier during the  
phase of the bounce. This potential, $\mathcal{V}(\eta) \equiv -m_{\rm
  eff}^2(\eta)$ in each case, can be, during the bounce phase,  approximated by a
P\"{o}schl-Teller potential,
\begin{equation}
\mathcal{V}_{PT}(\eta)= \mathcal{V}_{0}
\mathrm{cosh}^{-2}[\alpha(\eta-\eta_{B})],
\end{equation}
 for which we know the analytical solution.  In the latter equation, $\mathcal{V}_0$ is
 the  effective potential's height while $-2\mathcal{V}_{0}\alpha^2$ is the curvature of the potential at the
 maximum point . In the  dressed metric approach, the height
 $\mathcal{V}_0$ can be obtained from the expression of $a''/a$, being
 equal to $\mathcal{V}_0=k_{B}=\alpha^{2}/6$, while for the hybrid,
 the scale $k_B$ is replaced by $k_H$. Hereafter, we use the notation
 $k_{B/H}$ when we want to  refer  to the characteristic scale at the
 bounce in the dressed ($k_{B}$) and hybrid approaches ($k_{H}$),
 respectively.

The solution for $\chi_{k}$,  in the dressed and hybrid
approaches, can be written in the form of the standard
hypergeometric equation's solution,  given by~\cite{Zhu:2017jew}
\begin{eqnarray}
\!\!\!\!\!\!\!\!\!\!\!\!\!\! \chi_{k}(\eta) &=& a_{k}
x^{ik/2\alpha}(1-x)^{-ik/2\alpha}   \nonumber \\ &\times& _{2}F_{1}
(a_{1} - a_{3} +1, a_{2}-a_{3}+1, 2 -a_{3}, x)  \nonumber \\ &+& \;
b_{k} [x(1-x)]^{-ik/2\alpha} \;  _{2}F_{1}(a_{1}, a_{2}, a_{3}, x),
\label{bouncesol}
\end{eqnarray}
where $x\equiv x(\eta)=\{1+\exp[-2\alpha \, (\eta-\eta_{B})]\}^{-1}$,
\begin{eqnarray}
&&a_{1} \equiv
  \frac{1}{2}\left(1+\sqrt{1+\frac{32\pi\rho_c}{3\alpha^2}}\right) -
  \frac{ik}{\alpha},  \\ &&a_{2} \equiv
  \frac{1}{2}\left(1-\sqrt{1+\frac{32\pi\rho_c}{3\alpha^2}}\right) -
  \frac{ik}{\alpha},  \\ &&a_{3} \equiv 1 - \frac{ik}{\alpha},
\end{eqnarray}
and $a_{k}$ and $b_{k}$ are integration constants to be determined by
the initial conditions.

\subsection{Closed/Deformed Algebra Approach}

Within the framework of closed/deformed algebra approach, the
effective equation of motion for the  modes in  {}Fourier space is
given by~\cite{Bojowald:2008gz,Cailleteau:2011kr,Cailleteau:2012fy}
\begin{equation}\label{muclosed}
\chi_{k}''(\eta) +  W_{k, {\rm eff}}^2(\eta) \chi_{k}(\eta)=0,
\end{equation}
where
\begin{eqnarray}
\label{omegaCA}
 W_{k, {\rm
     eff}}^2(\eta)&=&\Omega(\eta)k^{2}-\frac{z''(\eta)}{z(\eta)},\\
\label{OmegaCA}
\Omega&\equiv& 1-\frac{2\rho}{\rho_c}.
\end{eqnarray}
When $W_{k, {\rm eff}}^2<0$, the modes are outside the Hubble horizon
and are decaying/growing modes, whereas for $W_{k, {\rm eff}}^2>0$ the
modes are inside the Hubble horizon and are oscillatory. 

In this approach, there is no effective potential $U(\eta)$ like in
the dressed metric and hybrid ones.  However, we need to be careful
with the factor $\Omega$, which change signs at $\rho=\rho_c/2$.  The
instant $t=t_S$ when $\rho=\rho_c/2$, i.e., $\Omega = 0$, is the so called
silent point. At $t_S$ all the space points are
uncorrelated~\cite{Mielczarek:2014kea}.  Depending on the signature of
$\Omega$, there can be two different regions, the Euclidean region
($\rho_c/2<\rho<\rho_c$) and the Lorentzian region ($\rho<\rho_c/2$).
In order to avoid difficulties in the calculations in the Euclidean
regime~\cite{Linsefors:2012et}, it is usual to consider the modes only
in the Lorentzian region, which means that $t\geq t_S$. 

\subsection{Gravitational Particle Production}

Let us now consider the evolution of the {}Fourier modes $\chi_k$ of
the (here assumed) massless spectator scalar field $\chi$, whose
equation of motion represent a set of uncoupled oscillators with a
  frequency which varies in time.  Due to the time variable frequency, we can define
a different vacuum for each instant $\eta$.  The effect of GPP was
introduced by Parker~\cite{Parker:1968mv,Parker:1969au}, who
developed an understanding about the  conditions for the definition of a  particle
number  $n(\eta)$ which is time dependent,  which are (i) its vacuum expectation value
varies sufficiently slowly  with time as the 
Universe expansion rate is enough slow and (ii) the  period of expansion must  
occur between the limit of two (``Minkowskian") vacuum states.    In the following
we summarized the mathematical treatment of GPP. 

The Hamiltonian for $\chi_k(\eta)$ can be written as follows:
\begin{eqnarray}\label{H}
\!\!\!\!\!\!\!\!\!\!\!\!\!\!\!\!H(\eta) \!=\! \int d^3k  \left(  2 E_k
\hat{a}_{\vec{k}}^\dagger \hat{a}_{\vec{k}} + F_{\vec{k}}
\hat{a}_{\vec{k}} \hat{a}_{-\vec{k}} + F_{\vec{k}}^*
\hat{a}_{\vec{k}}^\dagger \hat{a}_{-\vec{k}}^\dagger \right),
\end{eqnarray}
where
\begin{eqnarray}
&&F_k(\eta)=\frac{1}{2}(\chi_k'(\eta))^2+ \frac{\omega_k^2}{2}
(\chi_k(\eta))^2,
\label{Fw}
\\ 
 &&E_k(\eta)= \frac{1}{2}|\chi_k'(\eta)|^2 +
  \frac{\omega_k^2}{2}|\chi_k(\eta)|^2,
\label{Ew}
\end{eqnarray}
and $\omega_k(\eta)$ is the frequency in the  approximation
of a massless $\chi$ field.  The diagonalized Hamiltonian is obtained by
performing the  Bogoliubov transformation shown in the following equation:
\begin{eqnarray}
\hat{b}_{\vec{k}}=\alpha_k(\eta)\hat{a}_{\vec{k}} +\beta_k^*(\eta)
    {\hat{a}_{-\vec{k}}}^\dagger
\end{eqnarray}
where   $\beta_k(\eta)$ and $\alpha_k(\eta)$ 
satisfy the normalization constraint given by $ |\alpha_k(\eta)|^2  - |\beta_k(\eta)|^2=1$.  The diagonalized Hamiltonian
can be written as,
\begin{eqnarray}\label{Hdiag}
H(\eta)= \int d^3k \  \omega_k\  b_{\vec{k}}^\dagger\  b_{\vec{k}}.
\end{eqnarray}
Equation~(\ref{Ew})  can then be written as follows,
\begin{eqnarray}\label{Ek}
E_k(\eta)=\omega_k\left[ \frac{1}{2} + |\beta_k(\eta)|^2\right].
\end{eqnarray}  
We can define the vacuum states $|0_{(a)}\rangle$  and $|0_{(b)}\rangle$
in a way that $a_{\vec{k}}|0_{(a)}\rangle=b_{\vec{k}}|0_{(b)}\rangle=0$.
One can compute  the  number operator $\hat{N}_{\vec{k}}^{(b)}=b_{\vec{k}}^\dagger\  b_{\vec{k}}$ expectation value
 in the
vacuum $|0_{(a)}\rangle$, as given by 
\begin{eqnarray}\label{N}
n_k(\eta)=
\langle_{(a)}0|\hat{N}_{\vec{k}}^{(b)}|0_{(a)}\rangle=|\beta_k(\eta)|^2.
\end{eqnarray}
The quantity  $|\beta_k(\eta)|^2$ is interpreted as the particle
number per mode.  
The
initial Minkowski vacuum states $\chi_k^{(i)}$ can be related to the ones at a later time $\chi_k^{(f)}$  through the Bogoliubov coefficients:
\begin{eqnarray}\label{modesbogoliubov}
\chi_k^{(f)}(\eta)=\alpha_k \chi_k^{(i)}(\eta) + \beta_k
\chi_k^{(i)*}(\eta).
\end{eqnarray}
When  there are zero produced particles $\beta_k=0$ and
the normalization constraint implies 
$\alpha_k=1$ and $\chi_k^{(i)}=\chi_k^{(f)}$.

With the above expressions we can compute $n_p(\eta)$, the total particle number density   and $\rho_p(\eta)$, the
 energy density  gravitationally produced.  The 
 particle number density, which is integrated over all modes, is given by
\begin{eqnarray}\label{nv0}
n_p(\eta)&=&\frac{1}{a^3(\eta)
  L^3}\left(\frac{L}{2\pi}\right)^3\int\limits_0^\infty
d^3k\   n_k(\eta)  \nonumber
\\ &=&\frac{1}{2\pi^2a^3(\eta)}\int\limits_0^\infty dk \ k^2
|\beta_k(\eta)|^2,
\end{eqnarray}
whereas $\rho_p(\eta)$, the  energy density associated to the produced particles is given by
\begin{eqnarray}\label{rhov0}
\rho_p(\eta)=\frac{1}{2\pi^2a^4(\eta)}\int\limits_0^\infty dk\,
k^2\,\omega_k  |\beta_k(\eta)|^2.
\end{eqnarray} 
Equations~\eqref{nv0} and~\eqref{rhov0} give us the particle number
density and the energy density of the produced particles, respectively. 

It is important to mention that Eq.~(\ref{rhov0}) gives the net energy
density produced between two Minkowskian vacuum states.  On the other
hand, we can also obtain the energy density of produced particles due
to the GPP effect from the expectation value of the test field's energy-momentum
tensor   at a time $\eta$, that corresponds to~\cite{Bunch:1980vc}
\begin{equation}\label{rhoEM}
\rho_p^{\mathrm{EM}}(\eta)=\frac{1}{4\pi^2a^4(\eta)}\int\limits_0^\infty
dk\ k^2 \left[\frac{}{}|\chi_k'(\eta)|^2 + \omega_k^2|\chi_k(\eta)|^2
  \right].
\end{equation} 
%

The Minkowskian initial condition set at the contracting phase is
called the Bunch-Davies (BD) vacuum~\cite{Bunch:1978yq}.
Alternatively, it is also possible to impose an initial condition at
the bounce, which is the fourth-order adiabatic vacuum
state~\cite{Ashtekar:2009mm}.  However, it is important to  note that the
quantum contributions computed with a Bunch-Davies vacuum initial condition
and also with  fourth-order adiabatic vacuum  are the same just in the case of 
modes with $k \geq k_B$ ($k \geq k_H$) in the dressed metric (hybrid
approach).  {}For any other modes, the fourth-order adiabatic vacuum state
at the bounce may not be applicable~\cite{Zhu:2017jew}, while the Bunch-Davies
vacuum  can still be considered  in the
contracting phase.

We can also mention other two types of initial conditions, which are
the non oscillating
vacuum~\cite{deBlas:2016puz,ElizagaNavascues:2020fai,Gomar:2017yww}
and the silent point vacuum~\cite{Mielczarek:2014kea}. The former
relates to a method for minimizing the oscillations in the resulting
power spectrum of perturbations, which can be considered in dressed
and hybrid approaches, whereas the latter is particular to the
closed/deformed algebra approach and it is necessary for its
consistency.

Explicitly numerically solving for the mode equations given above is 
computationally intensive. This is particularly true in the regimes with
rapidly oscillating high-momentum modes. We need to also handle the UV divergences
that appear and then the particle production energy density needs
to be renormalized appropriately. A typical approach is to use a
Wentzel-Kramers-Brillouin (WKB) approximation for the modes to tackle 
these problems~\cite{Calzetta:2008iqa}. But even so,
there are issues with both how to fix the upper limit for the momentum
integrals and further issues in the infrared, which also demands
to consider a lower limit for the momentum integrals when computing
the total energy density due to GPP. This is also not free from ambiguities.
It is important, thus, to have a computation as analytical as possible and in such
ways one can overcome the above mentioned issues but still having a reliable
computation for the GPP. 
In the following section, we give approximate analytical solutions for
the equations of motion for the scalar modes in each approach and from which
we can estimate the GPP in appropriate ways.

\section{Results}
\label{sec4}

We present approximated analytical results for GPP in the dressed
metric, hybrid and closed/deformed algebra approaches of LQC.  To
obtain these results, instead of fully computing the real-time
backreaction  of the produced particles  in the background, we
 estimate the energy density associated to those  after the bounce,
and then compare the result with the background energy density.  As we
are going to see, these estimates are already sufficient for a
qualitative analysis. Later, we compare how the energy density  of the
produced particles redshifts with  the scale factor in comparison with
the behavior of the  dominant background energy content.  These
results will help to gauge the validity of each of those approaches in
LQC.  We compute the energy density of gravitationally produced
particles in each approach, which consists in computing the
corresponding analytical expression for $\beta_k$.

\subsubsection{Dressed Metric Approach}

{}From the results of Ref.~\cite{Zhu:2017jew}, by matching  the
analytical solutions for the bounce phase within the P\"{o}schl-Teller
potential approximation, the transition and  slow-roll phase,
the Bogoliubov coefficients can be obtained and they are given by
\begin{eqnarray}
\alpha_k  &=&  \sqrt{2k} \left[ a_k
  \frac{\Gamma(2-a_3)\Gamma(a_1+a_2-a_3)}{\Gamma(a_1-a_3+1)
    \Gamma(a_2-a_3+1)}\right.\nonumber\\ &&\left.+
  b_k \frac{\Gamma(a_3)\Gamma(a_1+a_2-a_3)}{\Gamma(a_1)\Gamma(a_2)}
  \right]e^{ik\eta_B},\\ \beta_k \label{betadressed} &=&  \sqrt{2k}
\left[ a_k
  \frac{\Gamma(2-a_3)\Gamma(a_3-a_1-a_2)}{\Gamma(1-a_1)\Gamma(1-a_2)}
  \right.\nonumber\\ &&\left. +b_k
  \frac{\Gamma(a_3)\Gamma(a_3-a_1-a_2)}{\Gamma(a3-a_1)\Gamma(a3-a_2)}
  \right]e^{-ik\eta_B},
\end{eqnarray}
where the pairs ($\alpha_k$, $\beta_k$) and ($a_k$, $b_k$) are
arbitrary constants at the bounce phase and slow-roll inflation
solution for the $k$- modes, respectively, and  $\eta_B$ is the
conformal time at the bounce.  The Bogoliubov coefficients are
determined when we impose initial conditions, i.e., choose $a_k$ and
$b_k$.

Assuming the absence of particles at the onset of inflation, one would
impose that  $\alpha_k=1$ and $\beta_k=0$.  However the value of
$\alpha_k$ and $\beta_k$ must be obtained starting from vacuum initial
conditions in the previous phases. Two different types of initial
conditions were already considered in the
literature~\cite{Zhu:2017jew}, which are the aforementioned BD vacuum
in the contracting phase~\cite{Barrau:2016nwy} and the fourth-order
adiabatic vacuum at the bounce~\cite{Agullo:2012fc}, which are,
respectively, given by 
\begin{eqnarray}\label{ics_dressed}
  \!\!\!\!\!\!\!\!\!\!\!\!\!\!\chi_k^{\mathrm{(BD)}}(\eta)
  &=&\frac{1}{2k}e^{-ik\eta},\label{ICmuBD}
  \\ \!\!\!\!\!\!\!\!\!\!\!\!\!\!\chi_k^{\mathrm{(WKB)}}(\eta)
  &=&\frac{1}{2k}\left[1-\frac{1}{4}\frac{k_B^2}{k^2}
    -\frac{29}{32}\frac{k_B^4}{k^4}
    +\mathcal{O}\left(\frac{k_B}{k}\right)^6\right],\label{ICmuWKB}
\end{eqnarray}
where $k_{B}=\sqrt{\gamma_{B}/3}\, a_{B} m_{Pl}$ is the 
energy scale in the bounce in the dressed approach. The ``WKB" in
Eq.~\eqref{ICmuWKB} refers to the WKB approximation, used to obtain
this result.  These initial conditions lead to the same results for
GPP in the case considered here and as computed explicitly in
Ref.~\cite{Graef:2020qwe}.  Setting the previous initial conditions,
it then follows from Eq.~\eqref{betadressed} that 
\begin{equation}\label{betadressed1}
|\beta_{k}|^{2} =
\frac{1}{2}\left[1+\cos\left(\frac{\pi}{\sqrt{3}}\right)\right] {\rm
  csch}^{2}\left(\frac{\pi k}{\sqrt{6}k_{B}}\right).
\end{equation}
Above $|\beta_{k}|^{2}$ corresponds to the number of particles per
mode $k$ that were produced, namely  $n_{k}$. Using this quantity, in
Ref.~\cite{Graef:2020qwe}  the energy density of particles produced
with and without backreaction was obtained.  However, unlike the procedure used in
that reference, here we are going to consider the contribution from
all the modes for the density of produced particles,  not  only  the
modes that in the pre-inflationary phase exit and then reenter the  effective horizon.  Note also that using all the modes is
typically the procedure adopted in studies of GPP in
general~\cite{Calzetta:2008iqa}.  The produced modes are effectively
considered as particles after they reenter the horizon. Therefore, the
energy density stored in the produced particles is then given by
\begin{eqnarray}
\label{rhofinal}
\rho_p(\eta)=\frac{1}{2\pi^2a^4(\eta)}\int_{0}^{\infty} dk\, k^2\,
n_k(\eta)\, \omega_k,
\end{eqnarray} 
We then obtain that
\begin{equation}\label{intfromzero}
\rho_{p}(\eta) = \frac{1+\cos\left(\frac{\pi}{\sqrt{3}}\right) }{4
  \pi^{2} a^{4}(\eta)} \int_{0}^{\infty} dk\, k^{3}\, {\rm
  csch}^{2}\left(\frac{\pi k}{\sqrt{6}k_{B}}\right),
\end{equation}
where we used  $n_k \equiv |\beta_k|^2$, where $|\beta_k|^2$ is given by
Eq.~\eqref{betadressed1} and $\omega_k\sim k$ in the case of relativistic 
particles.
By performing the integration, Eq.~(\ref{intfromzero}) gives 
\begin{equation}\label{rhoGPP}
\rho_{p}(\eta) \simeq 12.5\times 10^{-3} \frac{k_{B}^{4}}{a^{4}(\eta)}
\simeq 1.3 \frac{ m_{\rm Pl}^4}{a^{4}(\eta)}.
\end{equation}

We must remember that   the
background energy evolves dynamically as stiff matter (since the kinetic energy of the inflaton dominates in the bounce), with
$\rho_{bg}=\rho_{c}a^{-6}$. However the gravitationally
produced particles evolve like relativistic matter, with
$\rho_{p}\propto  a^{-4}$.  In a previous work \cite{Graef:2020qwe},
it was estimated that, in order to remain subdominant before and at the
beginning of inflation, the energy density on the produced
particles must satisfy the condition $\rho(t_s)< 2 \times
10^{-5}m_{\rm Pl}^{4}$. The time $t_s$, when $a''/a = 0$, is when the
maximum of GPP happens. This value was estimated by considering an
equal amount of energy density in radiation and in the inflaton
potential energy at the beginning of inflation, and then receding this
radiation density backward in time by multiplying it to $a^{4}$ until
the time $t=t_{s}$. By assuming $a_B=1$, at the time  $t_{s}= 0.3
t_{Pl}$, the scale factor is found to be $a(t_{s}) \approx 1.248$.
Thus, from Eq.~\eqref{rhoGPP} we can estimate that $\rho_{p}(t_s)
\approx 0.54 \; m_{\rm Pl}^{4}$. Therefore,  we confirm that, in the
context of the dressed metric approach, the GPP density for
relativistic $\chi$ particles will eventually dominate the dynamics,
which is inconsistent with the premise that backreaction must be small
for the dressed metric approach to be valid and as far as the
production of massless spectator scalar particles are concerned. 

\subsubsection{Hybrid Approach}

The analytic form of the solutions and the matching conditions in the
hybrid approach are analogous to the previous case (see
Ref.~\cite{Wu:2018sbr} for more details). It is straightforward  to
obtain  the Bogoliubov coefficients $\alpha_k$ and $\beta_k$ by
matching the solutions in  the bounce phase, transition phase and
slow-roll phase, whose procedure follows similarly to that done for
the dressed metric approach.

Considering the BD vacuum as the initial
condition~\cite{Martineau:2017sti} the Bogoliubov coefficients then
reads now
\begin{equation}\label{betaH}
|\beta_{k}|^{2} =
\frac{1}{2}\left[1+\cos\left(\frac{\sqrt{5}\pi}{\sqrt{3}}\right)\right]
     {\rm csch}^{2}\left(\frac{\pi k}{\sqrt{6}k_{H}}\right),
\end{equation}
where, as already defined earlier, $k_{H}=k_{B}/\sqrt{3}$ is the
 energy scale at the bounce in the hybrid approach.  The
quantum effects in both dressed metric and hybrid approaches effective
equations are qualitatively the same~\cite{Wu:2018sbr} but exhibit
two quantitative differences.  These differences are the
characteristic energy scales $k_B$ and $k_H$ and the numerical
factor before the hyperbolic function in Eqs.~\eqref{betadressed1}
and~\eqref{betaH}. In addition,   in the hybrid case we have a
positive time-dependent effective mass as one approaches the bounce,
while in the dressed metric case the time-dependent effective mass is
negative when approaching the bounce and around
it~\cite{ElizagaNavascues:2017avq}.

Quantitatively, by comparing Eq.~\eqref{betadressed1} with
Eq.~\eqref{betaH},  we notice that the only differences will be a
factor of $1/\sqrt{3}$ from the characteristic scale $k_{H}$ as
compared to $k_{B}$, in addition to the factor of $\sqrt{5}$ in the
cosine argument. Then, the corresponding expression to
Eq.~\eqref{intfromzero} in the hybrid approach is simply
\begin{equation}\label{intfromzeroH}
\rho_{p}(\eta) = \frac{1+\cos\left(\frac{\sqrt{5}\pi}{\sqrt{3}}\right)
}{4 \pi^{2} a^{4}(\eta)} \int_{0}^{\infty} dk\, k^{3}\, {\rm
  csch}^{2}\left(\frac{\pi k}{\sqrt{6}k_{H}}\right).
\end{equation}
Again, the integral is ultraviolet dominated and
Eq.~(\ref{intfromzeroH}) gives 
\begin{equation}\label{rhoGPPH}
\rho_{p}(\eta) \simeq 6.5\times 10^{-3} \frac{k_{H}^{4}}{a^{4}(\eta)}
\simeq 7. 5 \times 10^{-2} \frac{ m_{\rm Pl}^4}{a^{4}(\eta)}.
\end{equation}
If we estimate the energy density stored in the particles produced
only for the modes that  exit and reenter the effective horizon
$\lambda$ during the pre-inflationary phase we  obtain the value
$\rho_{p}(\eta)=10^{-3} m_{\rm Pl}^4/a^{4}(\eta)$.
Equation~\eqref{rhoGPPH} is only defined for $t \gtrsim t_s$, when the
modes are well inside the horizon to be considered as particles and
which is the moment when $\rho_{p}$ assumes its highest value. {}From
Eq.~\eqref{rhoGPPH}, we obtain that  $\rho(t_s) \approx 3.1\times
10^{-2} m_{\rm Pl}^4$ for the density of particles produced at
$t=t_{s}$, which is approximately the instant that GPP ceases.  As
discussed previously,  in order to remain subdominant until the start of  inflation, the  density  in the produced
particles must satisfy the condition $\rho_{s}< 2 \times 10^{-5}m_{\rm
  Pl}^{4}$. Despite having shown through Eq.~\eqref{rhoGPPH} that
$\rho_{s}$ is  smaller than the corresponding quantity in the dressed
metric approach, we can see that  the condition required for the
produced particles not to dominate the background energy density is
still {\it not} satisfied in the hybrid approach as well. Analogously
to what happens in the dressed approach, this invalidates the
effective description usually considered in these approaches. as far
as the production of massless spectator scalar particles are again
involved. 

Therefore, despite the difference in the maximum density of produced
particles, the conclusions for the dressed and hybrid approaches will be
basically the same.  In Ref.~\cite{deBlas:2016puz} (and more recently
in Ref.~\cite{ElizagaNavascues:2020fai}), another proposal was
suggested to select the initial vacuum state. It suggests to select the initial conditions for each mode in order to 
minimize the time variation of the spectator field amplitude since  the bounce until the starting of the inflationary phase. As shown in
Ref.~\cite{Gomar:2017yww},  such ``non-oscillating" initial conditions
lead to a primordial power spectrum without the large oscillations.  However, one can check that the number density of particles
produced in this case, given by the quantity $|\beta_{k}|^{2}$, will
not change considerably in our framework, since they get rid of the
oscillations by avoiding the fast oscillating term which we have
already averaged out in our analysis above. Therefore, one should not
expect that this method would  prevent excessive particle
production. This motivates us to move further to the investigation of
the particle production  in another framework. In the next section we
are going to consider  the closed/deformed algebra
approach, which  consists in another way of treating the
perturbations, possibly leading to different results. 

\subsubsection{Closed/Deformed Algebra Approach}

Here we follow the same approach considered in the previous cases to
obtain the parameter $\beta_k$ in the closed algebra
approach. However, this case is rather more involved than the previous
ones.  This is due to the fact that the description of the propagation
of the modes in the transition from the Lorentzian to the Euclidean
phase is not so rigorous~\cite{Mielczarek:2014kea} due to the presence
of the silent point.  This is called the {\it signature change
  problem}~\cite{Linsefors:2012et,Bolliet:2015raa,Bolliet:2015bka,Grain:2016jlq}.
The solution to this problem can be imposing initial conditions for
the modes at the silent point ($t=t_s$) in the Lorentzian phase after
the bounce, where the signature changes from Euclidean to Lorentzian.
In the silent point all points become uncorrelated, since the
space-dependent term in the equation of motion for the modes drops out
and the two-point function in this surface becomes zero.  Therefore,
after the silent point, in this approach, the  modes $\chi_k(\eta)$
obey Eq.~\eqref{muclosed}.  In particular, in the bounce and
transition phases, using the analytical approximations given by
Eqs.~\eqref{aLQCstiff} and~\eqref{phiLQCstiff}, from
Eq.~\eqref{OmegaCA}  we obtain that
\begin{eqnarray}\label{OmegaClosedapprox}
\Omega(\eta)&=&\frac{\tau^2-1}{\tau^2+1},
\end{eqnarray}
where $\tau=t/\tau_B$~\cite{Li:2018vzr}. 

We can obtain leading-order approximate solutions for the mode
functions of Eq.~\eqref{muclosed}  using the uniform asymptotic
approximation method~\cite{Zhu:2014aea,Zhu:2016srz}. The
complete evaluation of the solutions was presented in
Ref.~\cite{Li:2018vzr}. Here, we summarize the main steps for
completeness.  {}First, by appropriately changing variables, the mode
equation can be expressed in the form
\begin{equation}\label{chis}
\frac{d^{2}\chi_{k}(y)}{d y^{2}} =\left[ g(y)+q(y) \right]\chi_{k}(y),
\end{equation}
where $y=-k\eta$ and  the functions $g$  and $q$ are defined as
\begin{eqnarray}
\!\!\!\!\!\!g(y)&=&\frac{1-\tau^{2}}{1+\tau^{2}}+\frac{\gamma_B}{4k^2
  \tau^2}+\frac{\gamma_{B}(18+21\tau^{2}
  -\tau^{4})}{9k^{2}\tau^{2}(\tau^{2}+1)^{5/3}},\\ \!\!\!\!\!\!
q(y)&=&-\frac{\gamma_B}{4k^2
  \tau^2}.   
\end{eqnarray}
The analytical solution of Eq.~(\ref{chis}) is found to be given
by~\cite{Li:2018vzr}
\begin{eqnarray}\label{xiS}
\chi_k(t) = \left(\frac{\xi}{g}\right)^{1/4}
\left[a_k\mathrm{Ai}(\xi)+b_k\mathrm{Bi}(\xi)\right],
\end{eqnarray}
where $\mathrm{Ai}(\xi)$ and $\mathrm{Bi}(\xi)$ are the Airy functions
of the first and second kind, respectively, and the parameter $\xi$ is
related to the function $g$ through
\begin{eqnarray}
\!\!\!\!\!\xi &=&  \left\{ 
\begin{array}{lll}
&\left(-\frac{3k}{2}\int_{t_+}^{t}\frac{\sqrt{g}}{a}dt\right)^{2/3} &
  \mbox{, $t <
    t_+$}\\ -&\left(\frac{3k}{2}\int_{t_+}^{t}\frac{\sqrt{-g}}{a}dt\right)^{2/3}
  & \mbox{, $t > t_+$}
\end{array} 
\right. 
\end{eqnarray}
$t_+$ being the turning point (where $g(y)=0$).  Also, the Wronskian
condition implies that
\begin{eqnarray}
&&\chi_k \dot{\chi}_k^*-\chi_k^* \dot{\chi}_k=i/a,\\ &&a_k b_k^*-a_k^*
  b_k=i\pi/k.
\end{eqnarray}

The qualitative general behavior of the function $g(y)$ keeps the same for all modes,
having only one turning point for any $k$ in the bouncing and transition phases. However, the precise location of the turning point $t_+$
depends on the co-moving wavenumber $k$.

Even though we cannot explicitly obtain the GPP in general from the
above equations,  we can still get a clear picture of GPP in the small
and long wavelength approximations.

\paragraph{\bf The small wavelenghts regime}

In the transition phase, $\xi(t)$ approaches to asymptotic negative
infinity. In this region, the Airy functions assume their asymptotic
form. Considering the previous definitions given above and together
with the equation for the modes given by Eq.~\eqref{xiS}, we can then
write the solution for the modes as
\begin{eqnarray}
\chi_k(t) &=&
\frac{1}{\sqrt{\pi}(-g)^{1/4}}\left\{a_{k}\cos
\left[\frac{2}{3}(-\xi)^{3/2}-\frac{\pi}{4}\right]\right.
\nonumber\\  &+&\left. b_{k}\sin\left[\frac{2}{3}(-\xi)^{3/2}
  -\frac{\pi}{4}\right]\right\}.
\label{chikt}
\end{eqnarray}
After some algebra, it is possible to show that Eq.~(\ref{chikt}) can
be put in the form
\begin{eqnarray}
\chi_k&=&\frac{e^{-i\pi/4}}{2\sqrt{\pi}}(a_k-ib_k)e^{i k
  (\eta-\eta_{\rm fB})}\nonumber\\ &+&(ia_k- b_k)e^{-i k
  (\eta-\eta_{\rm fB})}.
\label{sol1}
\end{eqnarray}

On the other hand, when the horizon goes to
negative infinity in the transition phase, the equation of motion for the modes becomes 
\begin{eqnarray}
\chi_k''+k^{2}\chi_k=0,
\end{eqnarray}
whose solution is
\begin{eqnarray}
\chi_k&=&\frac{1}{\sqrt{2k}}(\tilde{\alpha}_ke^{-i k \eta}+
\tilde{\beta}_{k}e^{i k \eta}).
\label{sol2}
\end{eqnarray}

By comparing Eqs.~(\ref{sol1}) and (\ref{sol2}), we can match the two
sets of integration constants, which allows us to obtain the
coefficients in the UV limit:
\begin{eqnarray}\label{alfabeta}
\alpha_k&=&\sqrt{\frac{k}{2\pi}}(i a_k-b_k)e^{i k \eta_{\rm
    fB}-i\pi/4},\\ \beta_k&=&\sqrt{\frac{k}{2\pi}}(a_k-i b_k)e^{-i k
  \eta_{\rm fB}-i\pi/4},
\end{eqnarray}
where
\begin{eqnarray}
\eta_{\rm fB}=\eta_f - \int_{\eta_+}^{\eta_f}\sqrt{-g(\eta)}\,d\eta.
\end{eqnarray}
The coefficients $a_k$ and $b_k$ are obtained by matching the power
spectrum to the one given by GR. This is possible because in the
regime that we are interested, the equation of motion for the
spectator field modes $\chi_k$  has the same behavior as the equation
of motion of the inflaton field and of the curvature perturbations,
which are  basically due to the inflaton fluctuations. Therefore, the
dynamics of the modes we are computing here must not present
divergences, since those would be translated to divergences in the
power spectrum. In order to define a behavior for the scalar modes
that can be consistent with the observations in the closed algebra
approach, we will interpret such modes analogously to the ones which
will enter in the expression for the power spectrum of the model. Its
possible to show that the only possible initial condition that allows
 compatibility of the power spectrum with the current CMB data is
given by (for details, see, e.g., Ref.~\cite{Li:2018vzr})
\begin{eqnarray}
a_k=\sqrt{\frac{\pi}{2k}}\qquad ,\quad b_k=-i\sqrt{\frac{\pi}{2k}}.
\end{eqnarray}
These coefficients lead to a spectrum equal to the classical GR result
in the observational window (the observed modes in CMB correspond to
the UV limit).  By inserting these coefficients in
Eq.~\eqref{alfabeta}, we obtain that
\begin{eqnarray}
\alpha_k=ie^{i(k \eta_{\rm fB}-\pi/4)}\qquad ,\quad \beta_k= 0. 
\end{eqnarray}
Any other initial condition implies a correction term (with respect
to GR spectrum) proportional to the wavenumber, which leads to a
divergent spectrum in the UV. We can see from the above expressions
that $|\alpha_{k}|^{2}=1$, which is consistent with the condition
$|\alpha_{k}|^{2} - |\beta_{k}|^{2}=1$. This corresponds exactly to
the classical case in GR. As the parameter $|\beta_{k}|^{2}=0$, this
implies no gravitational particle production in the UV limit of
this model.   Therefore, in this framework, from the UV modes with
such initial conditions (the only ones that do not produce
divergences), we see no particle production in any scenario that
provides a power spectrum consistent with the data. However, as
discussed in Ref.~\cite{Li:2018vzr}, in the Planckian UV regime,  new
ingredients are expected to take place, as modified dispersion
relations, for example, which could avoid possible divergences or even
change the behavior of the modes. In the absence of a definite model
for this regime, we are instead  going to focus on the case of IR
modes in order to check whether some considerably energy density can
be gravitationally produced in this regime.  Since in the UV regime
either we have no particle production or otherwise a new physics
would be coming into play,  the lack of information required to obtain
definite results from the UV regime motivates us to rediscuss the
possible initial conditions in the context of the IR modes. In the
following we discuss the behavior of such modes.

\paragraph{\bf Long wavelengths regime}

In the IR regime, $k< m_{\rm Pl}$, through the bounce and transition
phases the equation for the modes is found to have the solution
\begin{equation}\label{musIR}
\chi_k(\eta) = a_{k}\, z(\eta)+b_{k}\,z(\eta)\int_{\eta_*}^{\eta_{\rm
    end}}\frac{d\eta'}{z^2(\eta)}+\mathcal{O}(k^2),
\end{equation}
where $\eta_*$ denotes some particular reference time.  It can be
shown that this result leads to the following IR limit of the  power
spectrum~\cite{Bolliet:2015bka}:
\begin{equation}\label{PIR}
 P^{IR} \approx \frac{k^{3}}{2 \pi^2}\left| b_{k}
 \int_{\eta_*}^{\eta_{\rm end}}\frac{d\eta'}{z^2(\eta)}\right|^2.
 \end{equation}
Here, we are going to consider  initial conditions (i.e., the
expressions for $a_{k}$ and $b_{k}$) that lead to a spectrum in
agreement with what we observe.  Starting from Eq.~(\ref{PIR}), we can
consider  different approaches that set the initial conditions at the
vicinity of the silent point. The choice of the vicinity of the
silent point to set the initial conditions is justified in order to
avoid the problems that can happen due to the change of signature close to the
bounce.  The calculations in the infrared regards the fact that,  when
the term $k^2$ is neglected, 
there can be analytical solutions of the mode function equations. Let us first write a more general
parametrization to the coefficients:
\begin{eqnarray}\label{parameterization}
a_{k} \equiv a_{0} k^{n}, \; \; \; \; b_{k} \equiv b_{0} e^{-i\theta}
k^{l} , 
\end{eqnarray}
where $\theta$ is the relative phase between $a_{k}$ and
$b_{k}$. Besides, $a_{0}$ and $b_{0}$ are both positive and
independent of $k$, and  have dimensions of $m_{\rm Pl}^{-2n}$ and
$m_{\rm Pl}^{-2l}$, respectively.  The quantities $a_{k}$ and $b_{k}$
satisfy the Wronskian condition
\begin{eqnarray}\label{parameterization2}
a_{k} b_{k}^{*} - a_{k}^{*} b_{k} = i. 
\end{eqnarray}
Considering the parametrization given by Eq.~(\ref{parameterization}),
the above condition implies in  
\begin{eqnarray}\label{wronski2}
 2 a_{0} b_{0} \sin(\theta) = 1, \; \; \; \; n + l = 0.
\end{eqnarray}
Therefore, we can see that the only initial condition that can imply a scale invariant spectrum is
\begin{eqnarray}
 a_{k} \propto k^{3/2}, \; \; \; \; b_{k} \propto k^{-3/2},
\end{eqnarray}
and it implies scale invariance at any time until the end of 
inflation, since in the IR regime  the  term  proportional to
$b_{k}$ in Eq.~\eqref{musIR} will be dominant.  The term $b_{k}$
can be obtained by matching the solution for the modes in the
contracting and the bounce phase. It is straightforward to show that
$b_{k}=(3i/\sqrt{2})\lambda_{0}k^{-3/2}$  (see, e.g.,
Ref.~\cite{Bolliet:2015bka} for further details).

By comparing the expression given by Eq.~(\ref{PIR}) with the GR
spectrum we can identify that 
 \begin{align}\label{PIR3}
 P^{IR} &\approx \frac{k^3}{2\pi^2}
 |b_{k}|^2\left(\int_{\eta_*}^{\eta_{\rm
     end}}\frac{d\eta'}{z^2(\eta)}\right)^{2} = P_{GR} |\alpha_{k} +
 \beta_{k}|^{2}& \nonumber \\ &\approx 2.2 \times  10^{-9} |\alpha_{k}
 + \beta_{k}|^{2}.&
 \end{align}
Despite the initial conditions being related to the Bogoliubov
coefficients through Eq.~\eqref{PIR3}, it is not possible to identify
which contribution corresponds specifically to $\beta_k$, which enters
in the expression of the  particle production. However, the above
equation allows us to establish an upper limit in $\beta_{k}$.  Taking
into account the well known property $|\alpha_{k} + \beta_{k}|
\lesssim |\alpha_{k}| + |\beta_{k}|$ and remembering the normalization
condition  $|\alpha_k|^{2} - |\beta_k|^{2}=1$, we can show that 
 \begin{equation}
 |\alpha_{k} + \beta_{k}|  \lesssim |\alpha_{k}| + |\beta_{k}|
 \lesssim 1 + 2|\beta_{k}|.
 \end{equation}
 Therefore, from Eq.~(\ref{PIR3}) we obtain the limit
 \begin{equation}\label{limit}
 |\beta_k|^2  \lesssim  \frac{\lambda_{0}^{2} \times 10^{9}}{\pi^2}
 \left| \int_{\eta_*}^{\eta_{\rm
     end}}\frac{d\eta'}{z^2(\eta)}\right|^2,
 \end{equation}
where  $\eta_{\rm end}$ is the value of $\eta$ at the end of inflation
and $z(\eta)=a\dot{\phi}/H$. In the above equation, we made use of
Eq.~\eqref{PIR3} and that  $b_{k}=(3i/\sqrt{2})\lambda_{0}k^{-3/2}$.
The integral in Eq.~(\ref{limit}) depends only on the background,
being independent of $k$. At this point it is important to stress that,
while we have argued that it can be meaningless to define two-point
correlation functions before the silent point, concerning the
background there is in fact no problem in starting  the initial
conditions in the contracting phase. This was done in
Ref.~\cite{Li:2018vzr} and we will consider the same here. In this
case, for the background dynamics we can consider the evolution
presented in Sec.~\ref{sec2}.  During slow-roll inflation, when the
spectrum is computed, we have that $z(\eta)^{2}=2 a^{2}\epsilon$,
 $\epsilon$ being the slow-roll parameter given by $\epsilon
\simeq\dot{\phi}^{2}/(2H^{2})$ during inflation. Therefore, from now
on we can consider the approximation  $z(\eta)\approx
a(\eta)\sqrt{4\pi/(3m_{\rm Pl}^{2})}$.  Therefore, the integration in
Eq.~(\ref{limit})  can be computed using  Eq.~\eqref{HSR}. This
procedure leads in particular to the result
\begin{equation}
I(\eta_{e})\equiv \int_{\eta_*}^{\eta_{\rm
    end}}\frac{d\eta'}{a^2(\eta)} =  \frac{-m}{18\lambda_{0}}
\frac{1}{|\cos\theta_{A}|}\ln
\left(\frac{1}{2}\Gamma\sqrt{\frac{|cos\theta_{A}|}{f}}\right).
\label{integral}
 \end{equation}
 In the above expression we can use the values $m=10^{-6}m_{\rm Pl}$,
 $\Gamma=2\times 10^{-7}$, $f\approx 0.18$,  $\lambda_{0}=a_{in}
 H_{in}^2/4$   and $cos\, \theta_{A} \approx 1$, as obtained and
 discussed in Sec.~\ref{sec2}. One must remember that $\Gamma$ is the
 ratio between timescales as defined in Eq.~\eqref{Gammadef}, while
 the quantity $\Gamma \cos\, \theta_{A}$ corresponds to the value of
 $y$ at the onset of the bounce phase. The quantity $f$ is associated
 with the time of the beginning of the slow-roll phase, since
 $t_{SR}=t_{B}+f/m$ with $f$ defined in Eq.~\eqref{fdef}. With such
 values, we obtain that
 Eq.~(\ref{integral}) is estimated to be  $ I(\eta_{e}) \sim
 10^{-6}/\lambda_{0}$.  Therefore, using the expression for $b_{k}$,
 Eq. (\ref{limit}) becomes
 \begin{equation}\label{limit2}
 |\beta_k|^2 \lesssim \frac{\lambda_{0}^{2} \times 10^{9}}{\pi^2}
 \frac{I(\eta_{e})^{2}}{16} \approx 10^{-5}. 
 \end{equation}
 By substituting $|\beta_{k}|^{2}$ from the equation above  in
 Eq.~(\ref{rhofinal}), we obtain the following expression for the
 estimated upper limit on the density of particles gravitationally
 produced:
  \begin{eqnarray}\label{rho_closed}
 \rho_{p}  \lesssim \frac{ 10^{-5}}{2\pi^{2}a^{4}}\int^{m_{\rm
     Pl}}_{0} dk k^{3} \approx \frac{10^{-6}}{a^{4}} m_{\rm Pl}^{4},
 \end{eqnarray}
 where we only have integrated modes with $k \lesssim m_{\rm Pl}$
 since we are restricted to the IR limit.  In the latter equation, we
 have considered $\omega_{k}(\eta) \approx \sqrt{\Omega(\eta)} \; k$,
 since the modes only contribute to the density of particles after
 they are well inside the effective horizon. Also, we see  that the
 expression of $\omega_k$ has a correction factor in the closed
 algebra approach (see Eqs.~\eqref{omegaCA} and~\eqref{OmegaCA}),
 which is given by $\Omega(\eta)=1 - 2\rho/\rho_{c}$. In the silent
 point we have $\Omega=0$, and after that $\Omega(\eta)$ increases
 until it reaches the value $\Omega=1$, when $\rho\ll\rho_{c}$. Since
 $\Omega(\eta)$ does not depend on $k$, we simply consider its upper
 limit $\Omega=1$ in the above equation, since we want to obtain an
 upper limit for the density of produced particles. The fact that
 $\Omega$ is zero at the silent point and then increases means that in
 fact the particles start being produced right  after the silent
 point.  Since the particles produced behave as radiation, its energy
 density will then evolve as
  \begin{equation}
      \rho_{p}=\rho_{s}a^{-4}, 
  \end{equation}
where $\rho_{s}$ is the density of particles produced right after the
silent point. By comparing it with Eq.~(\ref{rho_closed}), we can see
that
  \begin{equation}\label{cond}
      \rho_{s} \lesssim 10^{-6} m_{\rm Pl}^{4},
  \end{equation}
which is right after the silent point, when the important modes are
already inside the horizon.  In order to know if  the produced
particles will not dominate the energy content of the Universe before
inflation, we must compare the density of particles produced,
$\rho_{p}$, with the background energy density $\rho_{bg}$ at the
beginning of inflation. Unlike the particles produced, which behave as
radiation, $\rho_{p}=\rho_{s}a^{-4}$, the background energy density,
on the other hand, evolves as  stiff matter,
$\rho_{bg}=\rho_{c}a^{-6}$, before inflation sets in.   Therefore, in
order for the density of produced particles not to come to dominate
before the onset of inflation, we must have the following  condition
satisfied:
 \begin{equation*}
     \rho_{s}a^{-4} < \rho_{c}a^{-6},
 \end{equation*}
and which must be satisfied in the beginning of inflation.  The value
of the scale factor in the beginning of inflation depends on the value
of some parameters associated with the initial conditions. However,
since the evolution of the background is the same in all approaches,
based on previous works (see, e.g., Refs.~\cite{Barboza:2020jux,Barboza:2022hng} for
example) we can estimate an amount of 4 to 5 e-folds of
pre-inflationary expansion (from the bounce to the beginning of
inflation). In the case of 4 pre-inflationary e-folds, we have that in
the onset of inflation $a_{\rm infl}\sim 55$. We can consider that
near the silent point $\rho_{s}=\rho_{c}/2\sim\rho_{c}a^{-6}_{s}$,
which leads to $a_{s}^{-6} \sim 1/2$ and, consequently, $a_{s}^{6}
\sim 2$. Therefore, it is a good approximation to consider
$a_{s}\approx 1$. Considering this value, the condition for the
produced particles not to come to dominate can be written as $\rho_{s}
\lesssim  10^{-4} m_{\rm Pl}^{4}$. Since Eq.~(\ref{cond}) shows that
$\rho_{s} \sim 10^{-6} m_{\rm Pl}^{4}$, we can safely conclude that
the particles gravitationally produced will not dominate the
background energy density before inflation in the deformed algebra
approach. This proves that the test field approximation is
consistently valid in this approach with this choice of initial
conditions, unlike in the hybrid and dressed approaches.

The violation of adiabaticity happens during a short time interval
around the bounce phase. 
This coincides with the phase when the main modes exit the effective horizon. 
These modes will then reenter the horizon after the bounce phase,  when 
they start  behaving as actual particles. 
Therefore, the backreaction is expected to be not strong enough to change our
conclusions.
This is also corroborated by a previous analysis  made in Ref.~\cite{Graef:2020qwe},
which was, however, restricted to the dressed case 
(for related work on the backreaction of GPP in general, see, e.g.,
Ref.~\cite{Tavakoli:2021kwj}).

\section{Conclusions} 
\label{sec5}

Given the importance of the effective description of LQC  in
providing means to obtain the relevant cosmological quantities, it is
of utmost importance to further analyze the validity of such a
description. Motivated by the results obtained in a previous
work~\cite{Graef:2020qwe}, we extended the analysis of the
backreaction from gravitational particle  production to other
approaches. 

{}Firstly, for the hybrid approach, we obtain a result similar to the
case of the dressed metric one, where the energy density stored in the
particles produced during the bounce phase dominates the energy
content of the Universe prior to inflation. Therefore,  if we extend
the validity of the effective description beyond the test field
approximation, this would imply a pre-inflationary radiation-dominated phase in these scenarios. 
This scenario would be similar to
a model of including the radiation effects in LQC, as studied in
Ref.~\cite{Barboza:2020jux}. A radiation-dominated phase in the
earlier stages of  expansion in these scenarios tends to imply a
small delay in the beginning of the inflationary phase in such
models. Also, the backreaction effect leads to a state significantly
different from the BD vacuum at the beginning of inflation. Indeed,
radiation has been shown to be an important factor in setting
 initial conditions for inflation appropriately (see, e.g.,
Ref.~\cite{Bastero-Gil:2016mrl} for a further discussion). However,
since one should not expect that the dressed metric and hybrid
approaches  could be consistent in such a regime, this analysis
actually put in check the validity of these approaches with the
initial conditions considered.

On the other hand, in the case of the closed algebra approach, we
obtain that the process of gravitational particle production leads to
a negligible backreaction effect. The energy stored in the produced
particles is very small compared to the energy density of the
background all the way up to the onset of inflation. This result was
obtained by considering initial conditions in the vicinity of the
silent point, which is justified in order to avoid  problems coming from a signature change close to  the bounce. Our result
corroborates the validity of the test field approximation in this
framework, showing  the robustness of the  effective description of
LQC in the closed algebra approach. Nevertheless we must point out
that this result is strictly related to the initial conditions which
are chosen in such a way that guarantees the consistency of the  model
with CMB data. Any dynamics that could lead to significant particle
production in this scenario would imply a divergent power spectrum.

In order to further  confirm the analytical results obtained here, it
would be important to perform a  numerical analysis capable of
including the backreaction effects of the particles in the background
simultaneously to its production. This will be done in a future work.


\section*{Acknowledgement}

The authors would like to thank A. Wang for the useful discussions.
 R.O.R. is partially supported by research
grants from CNPq, Grant No. 307286/2021-5, and from FAPERJ, Grant
No. E-26/201.150/2021. 
L.L.G is supported by Conselho Nacional de Desenvolvimento
Cient\'{\i}fico e Tecnol\'ogico (CNPq), under Grant
No. 307052/2019-2, and by the Funda\c{c}\~ao Carlos Chagas Filho de
Amparo \`a Pesquisa do Estado do Rio de Janeiro (FAPERJ), Grant
No. E-26/201.297/2021. L.L.G. would like to thank the Kavli Institute for the Physics and Mathematics of the Universe (IPMU) for the kind hospitality.


\end{document}